\documentclass[nofootinbib,twocolumn]{revtex4}

\usepackage{amssymb,amstext,amsmath}
\usepackage[dvips]{graphicx}
\usepackage{latexsym}
\usepackage{psfrag}
\usepackage{amsfonts}
\usepackage{bbm}
\usepackage{color}

\newcommand{\Ref}[1]{(\ref{#1})}

\newcommand{\eqa}{\begin{eqnarray}}
\newcommand{\neqa}{\end{eqnarray}}
\newcommand{\equ}{\begin{equation}}
\newcommand{\nequ}{\end{equation}}

\def\la{\langle}
\def\ra{\rangle}
\newcommand{\bra}[1]{\la {#1}|}
\newcommand{\ket}[1]{|{#1}\ra}

\def\d{\delta}
\def\f{\frac}

\newcommand{\p}{\partial}

\usepackage{bbm}

\newcommand{\SO}{\mathrm{SO}}

\newcommand{\lp}{\ell_{\rm P}}

\newcommand{\be}{\begin{equation}}
\newcommand{\ee}{\end{equation}}

\begin{document}

\title{\Large\bf Numerical evidence of regularized correlations in spin foam gravity}
\author{\large J. Daniel Christensen{}$^a$, Etera R. Livine{}$^b$ and Simone Speziale{}$^c$}

\affiliation{$^a$ Department of Mathematics, The University of Western Ontario, London, ON N6A 5B7, Canada.
\\ $^b$ Laboratoire de Physique, ENS Lyon, 
46 All\'ee d'Italie, 69364 Lyon, France.
\\ $^c$ Perimeter Institute, 31 Caroline St.\ N, Waterloo, ON N2L 2Y5, Canada.}

\begin{abstract}{
\noindent We report on the numerical analysis of the area correlations in spin foam gravity on 
a single 4-simplex considered by Rovelli in PRL {\bf 97} (2006) 151301. We compare the asymptotics and confirm the 
inverse squared distance leading behaviour at large scales. This supports the recent advances on testing the
semiclassical limit of the theory. Furthermore, we show that
the microscopic discreteness of the theory dynamically suppresses and regularizes the correlations
at the Planck scale.
}
\end{abstract}

\maketitle


A fundamental theory of quantum gravity is expected to improve the UV behaviour of
the non-renormalizable perturbative quantization of General Relativity. The latter should nonetheless
emerge in the low-energy limit, where it can be considered as an effective field theory.
Among the key features to be reproduced is the distance dependence of the free graviton propagator encoding Newton's law.
In Loop Quantum Gravity (LQG) and its covariant version, the spin foam formalism~\cite{book}, 
the UV behaviour is expected to be cured by the discreteness
of spacetime at the Planck scale. While the regularizing effect of such discreteness is clear, 
it is rather non-trivial how the discreteness smooths out to a low energy limit given by the effective
theory of gravitons. 
This is the problem of the semiclassical limit in LQG and spin foams.
In this letter we give numerical evidence that a behaviour consistent with the graviton theory at low energies
is dynamically suppressed and regularized at the Planck scale.

The study of the semiclassical limit has received a great deal of attention over the last few years
and important results have been obtained~\cite{cosmo}.
In particular, in~\cite{Modesto,RovelliProp,grav2,grav3,Io} a program was started to compute the graviton
propagator using correlations between geometric quantities. Analytic results show that the leading order
at large scales is consistent with that of the free propagator
from the linearized quantum theory, thus
providing an important piece of evidence for the correctness of the limit. 
Here we confirm this result numerically, supporting the approximations made
to deal with the complexity of the calculations.
Furthermore, the numerical approach also allows us to study the small scale structure of the propagator, 
where the non-perturbative effects of spin foams are dominant, and we show that
the discrete structure does regularize the typically divergent behaviour of the correlations, 
suppressing them at the Planck scale.
Finally, we point out the limitations of the model used, and discuss the relevant developments to be made.

We consider the area correlations for 4d Riemannian quantum gravity defined and studied in \cite{RovelliProp}. 
These are correlations between fluctuations of the areas around a given background $q$ of a 4-simplex, and
correspond to some components of the graviton propagator 
$G_{\mu\nu\rho\sigma}(x,y) = \bra{0} h_{\mu\nu}(x) h_{\rho\sigma}(y) \ket{0}$.
For extensive motivation and discussion, see \cite{grav2}. We index the ten triangles 
of the 4-simplex by $l$, and the area eigenvalues are given by $A_l=\lp^2 (2j_l+1)$, with $j_l$ half-integers.
For simplicity, we choose the background to be a regular four-simplex, with all ten areas having the same value, 
$A_0=\lp^2 (2j_0+1)$.
Given two triangles $a$ and $b$, we consider the following area correlation,
\equ\label{W}
W_{ab}(j_0)=\f1{\cal N} \sum_{j_l} 
\, {\mathbbm h}(j_a) \, {\mathbbm h}(j_b) \, \Psi_q[j_l] \, K[j_l],
\nequ
where ${\cal N}= \sum_{j_l} \Psi_q[j_l] \, K[j_l]$ is the normalisation,
$\Psi_q$ is the boundary state and $K$ the propagation kernel, or path integral amplitude.
These are model-dependent quantities, and we describe our choices below.
The quantity ${\mathbbm h}(j_a)\equiv \big(A_a^2-A_0^2\big)/A_0^2$ represents an area fluctuation,
or equivalently, the fluctuation of the metric tensor $h_{\mu\nu}$ projected along the normal
$n_a^\mu$ to the triangle. $W_{ab}(j_0)$ is the spin foam discrete analogue of the projections
$G_{ab} = n_a^\mu n_a^\nu n_b^\rho n_b^\sigma \, G_{\mu\nu\rho\sigma}(a,b)$
of the continuum graviton propagator around a flat background, with the two points taken to be the centers of the
triangles. If the theory has the right semiclassical limit, the leading order of \Ref{W} should match the leading order of $G_{ab}$, namely the free propagator corresponding to the linearized theory, which we recall scales as the inverse squared distance between $a$ and $b$.

The boundary state $\Psi_q[j_l]$ represents a dynamical coherent state peaked
around both the (canonically conjugate) intrinsic and extrinsic 3-geometry of the classical background $q$ chosen. 
Given the discrete 4-simplex considered, this means
choosing a background configuration for the areas and their conjugate variables, the dihedral angles. 
Taking the simple choice of the equilateral configuration $(j_0, \theta)$,
all triangle areas are given by $A_0=\lp^2(2j_0+1)$ and all inside dihedral angles by $\theta=\arccos(\f14)$.
The explicit form of such a state is not known in the full theory, but control can be 
gained by going to lowest order in the perturbative expansion. Based on analogies
with the continuum linearized theory, in \cite{RovelliProp} the following Gaussian ansatz was made,
\equ\label{Psi}
\Psi_q[j_l] = \exp\Big\{-\f1{2j_0} \sum_{lm} \alpha_{lm} \, \d j_l \, \d j_m + i \theta \sum_l (2j_l+1)\Big\},
\nequ
where $\alpha_{lm}$ is a 10 by 10 constant matrix and $\d j_l=j_l-j_0$. The matrix
$\alpha_{lm}$ is non-diagonal, but the symmetries of the equilateral background 
reduce the number of independent entries to three (see discussion below).
Having the squared width proportional to $j_0$ guarantees that in the large $j_0$ limit the state \Ref{Psi}
is peaked around both conjugate variables of the background $q=(j_0, \theta)$ (e.g. \cite{ClassTet}). 
The physical boundary state for the full theory is expected
to have a lowest order contribution, corresponding to the free theory, given by \Ref{Psi} with a
definite $\alpha_{lm}$. So if we know the full state, we can fix $\alpha_{lm}$ looking at its
perturbative expansion.
In this context, the parameter for the perturbative expansion can be taken 
to be $j_0$: as geometric areas are given by $\lp^2 \, {j_0}$, the limit $j_0\mapsto\infty$ drives the $\lp$ expansion.
Therefore, the parameter $j_0$ has a double role: on the one hand, it describes the background
geometry of the boundary; on the other hand, is the parameter of the perturbative expansion.

The kernel depends on the spin foam model chosen. As in \cite{RovelliProp},
we consider the Barrett-Crane (BC) model \cite{BC},
\equ\label{K}
K[j_l] = \prod_l (2{j_l}+1)^k  \, \{10j\},
\nequ
where the integer $k$ parametrizes the choice of face weight in the measure,
and the $10j$-symbol $\{10j\}$ is an $\SO(4)$ invariant tensor constructed with Clebsch-Gordan coefficients.
To study its perturbative expansion, recall that when all spins are homogeneously large,
i.e.\ $j_l = N k_l$, $N \mapsto \infty$, the $10j$-symbol has a stationary phase contribution
of the form \cite{asympt}
\equ\label{asymp}
 \mu(j_l) \cos \left(S_{\rm R}[j_l] + \phi\right),
\nequ
where $\mu(j_l)$ is a non-oscillating function scaling like $N^{-9/2}$,
$S_{\rm R}[j_l] = \sum_l (2j_l+1) \theta_l(j_l)$ is the Regge action for
a single 4-simplex with triangle areas $A_l = \lp^2(2j_l+1)$ as independent
variables, and $\phi$ is an irrelevant phase.
It was shown in \cite{asymptbaez} (see also \cite{asymptaltri}) that this is
masked by a non-oscillating contribution $D(j_l)$ that scales like $N^{-2}$.
This dominant contribution corresponds to a degenerate geometry for the 4-simplex.
However, as conjectured in \cite{RovelliProp} and proved in \cite{grav3},
$D(j_l)$ is negligible in the evaluation of quantities like \Ref{W}, where its
non-oscillating nature fails to properly match the phase of the boundary state \Ref{Psi}.
The work of this paper supports these analytic calculations, and thus
is the first (indirect) numerical evidence of \Ref{asymp}.
Therefore the asymptotic behaviour of \Ref{K} is effectively given by the kernel for Regge calculus
with measure $\prod_l (2{j_l}+1)^k \, \mu(j_l)$. 

The emergence of Regge calculus in the semiclassical limit of LQG
and spin foams has often been advocated \cite{Immirzi}, and would provide a solid bridge to low-energy physics: Regge calculus is a discrete representation of GR known to reproduce
the correct linearized quantum theory \cite{Ruth}, thus if in \Ref{W} we use $\Psi_q$ and $K$ from linearized
Regge calculus, we expect to obtain the right free graviton propagator.
This suggests a possible strategy to fix $\alpha_{lm}$ by evaluating the boundary state in linearized
Regge calculus \cite{Bianca}. 
In general, the matrix $\alpha_{lm}$ is non-diagonal, and this makes evaluating \Ref{W}
extremely challenging.  
To simplify the numerical analysis we study the case where $\alpha_{lm}\equiv \alpha \, \d_{lm}$ is diagonal.
Indeed, using the formula for the $\{10j\}$-symbol as an integral over $\SO(4)$, this 
choice allows us to 
perform each sum over $j_l$ separately, which simplifies the numerical task. The integral is
then dealt with using Monte Carlo methods.
We will come back to this point at the end, and discuss improvements in the boundary state. 

We then have a model with a single free parameter $\alpha$, determining the width of the Gaussian.
For different values of $\alpha$, we study the matrix \Ref{W} as a function of $j_0$,
in particular to test its asymptotic behaviour and support the analytic calculations which appeared
in \cite{RovelliProp}, which we briefly recall here.
First of all, $\prod_l (2{j_l}+1)^k \, \mu(j_l)$ is a measure term in \Ref{W}, thus it will affect 
only the higher order corrections \cite{Io}. The leading
order is obtained by approximating
\Ref{W} with (the second moment of) a Gaussian integral, like in the
continuum linearized theory, with 
action $Q_{lm} \, \d j_l \, \d j_m$ given by the quadratic term in the boundary state and the 
second derivatives of the Regge action,
\equ
Q_{lm} = \f{\alpha}{j_0} \, \d_{lm} + i \, \f{\p^2 S_{\rm R}}{\p j_l \, \p j_m}\Bigg|_{j_0}.
\nequ
The Hessian matrix of the Regge action on the equilateral configuration
was computed in \cite{grav2}, and it is a 10 by 10 matrix with a regular structure
inherited from the regularity of the equilateral 4-simplex. Fixing a triangle $a$, there are only three 
distinguishable choices for $b$: the case when $b$ is the same triangle, the six cases when $b$ is 
an adjacent triangle sharing an edge with $a$, and the three cases when $b$ is an opposite triangle
sharing only a vertex with $a$. Correspondingly, for each row of the Hessian there are only three different entries, 
\equ\label{H}
\f{\p^2 S_{\rm R}}{\p j_l \, \p j_m} = \f1{j_0} \, H_{lm},
\qquad H_{lm} = \f12 \sqrt{\f35} \, f_{lm},
\nequ
where for each row $f_{lm}=-9$ occurs once, $7/2$ six times, and $-4$ three times.
We can then write
\equ\label{G}
Q_{lm} = \f1{j_0} A_{lm}, \qquad A_{lm} \equiv \Big(\alpha \, \d_{lm} + i \, H_{lm}\Big).
\nequ
Notice that $H_{lm}$ is not invertible, just like the original continuum term, 
due to diffeomorphism invariance \cite{Bianca}. On the other hand,
$A_{lm}$ is invertible for any $\alpha>0$, thus the boundary state effectively 
provides a gauge-fixing for the propagator.

Following the same procedure as \cite{RovelliProp},
we obtain for the absolute value of the leading order
\equ\label{LO}
|W^{\rm LO}_{ab}(\alpha)| = \f4{j_0} \, |A^{-1}{}_{ab}|.
\nequ
If the theory has the right semiclassical limit, \Ref{LO} should give the free graviton propagator
of linearized quantum gravity.\footnote{We take the absolute value for a better comparison with the
linearized $G_{ab}$, because 
the spin foam kernel \Ref{asymp} provides the complex exponential of the Regge action even in
Riemannian signature.
The phase of \Ref{LO} is irrelevant for our work.} 
Choosing the harmonic gauge, the latter is given by
\equ\label{Wlin}
G_{\mu\nu\rho\sigma}(a,b) = -\f12
\f{\d_{\mu \rho}\d_{\nu\sigma}+\d_{\nu \rho}\d_{\mu\sigma}-\d_{\mu\nu}\d_{\rho\sigma}}{d(a,b)^2}.
\nequ
For the equilateral background we have chosen, there are only three independent projections $G_{ab}$,
and furthermore the squared distances are all proportional to $j_0$.
Both features are matched by \Ref{LO}, which has only the three
independent entries discussed above and scales as $1/j_0$.\footnote{One of the three cases is
the correlation of a triangle with itself. 
While the continuum graviton propagator between the same two points is divergent, this is not
the case for $W_{aa}$. This should not be surprising, as it is among the regularizing effects of the discrete
microscopic structure.} 

\begin{figure}[ht]
\includegraphics[width=8.3cm]{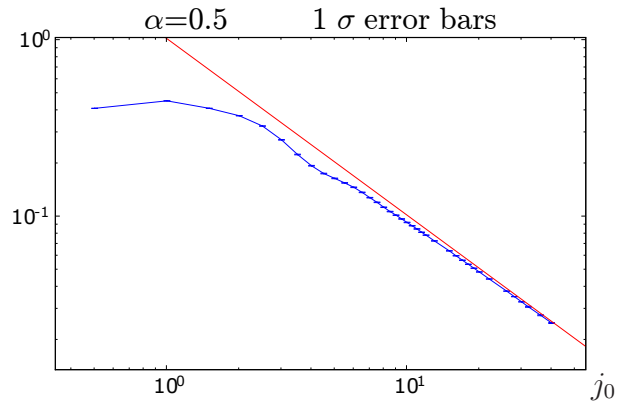}

\begin{picture}(0,0)(0,0)\put(110,12){\large$j_0$}\end{picture}
\vspace*{10pt}

\includegraphics[width=8.3cm]{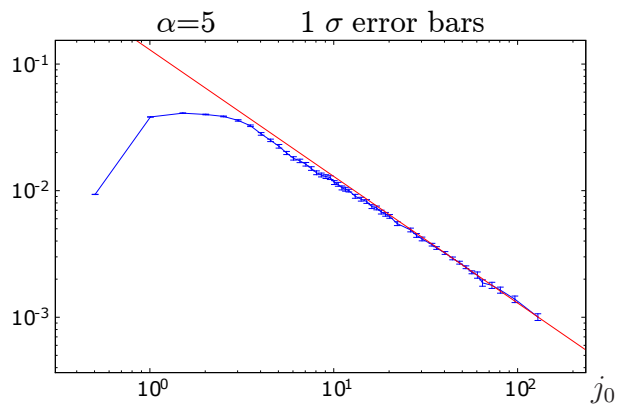}

\begin{picture}(0,0)(0,0)\put(110,12){\large$j_0$}\end{picture}
\vspace*{-10pt}
\caption{\small{Numerical study of \Ref{W} (dots), versus the analytic result
of the leading order, on a log-log plot.
Top panel: the case $\alpha=0.5$. Bottom panel: the case $\alpha=5$.
Raw data and more plots are available at
http://jdc.math.uwo.ca/graviton}}\label{fig}
\end{figure}

In general, $|W^{\rm LO}(\alpha)|$ scales as $(\alpha j_0)^{-1}$ for both
$\alpha\gg 1$ and $\alpha\ll 1$. To have explicit values, we choose
$a$ and $b$ to be opposite triangles, and fix $\alpha=0.5$ and $\alpha=5$. From \Ref{LO} we obtain 
\equ\label{LOnum}
|W^{\rm LO}_{\rm opp}(0.5)| = \f{1.02}{j_0}, \qquad
|W^{\rm LO}_{\rm opp}(5)| = \f{0.13}{j_0}.
\nequ
In Fig.~\ref{fig} we compare these analytic results with the numerical analysis of the full formula
\Ref{W}, where in the kernel \Ref{K} we choose the simple measure term with $k=0$.
The dots are the numerical evaluations of 
the absolute value of \Ref{W} for these cases, and the error bars are one $\sigma$.
Remarkably, good agreement is reached already at $j_0\sim 50$.
This is our first result: we numerically tested the inverse squared distance
asymptotic behaviour of the area correlation in spin foam gravity. This is crucial to support the approximations
used in \cite{RovelliProp} to deal with the complexity of the calculation, in particular the conjecture
that the $D$ term drops out.

\begin{figure}[ht]
\includegraphics[width=8.3cm]{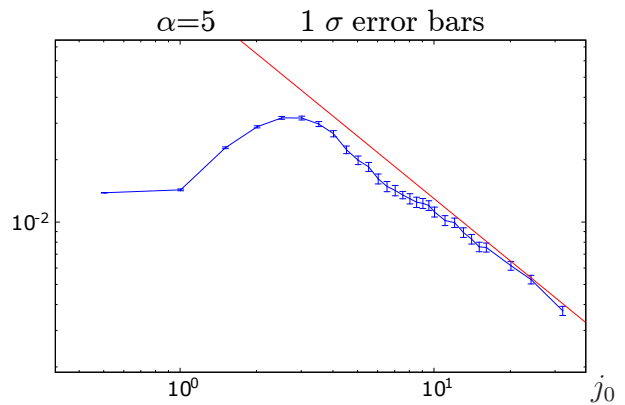}

\begin{picture}(0,0)(0,0)\put(110,12){\large$j_0$}\end{picture}
\vspace*{-10pt}
\caption{\small{The case $k=1$, $\alpha=5$. The leading order is the same as for $k=0$,
and the position of the peak shifted to the right.}}\label{kfig}
\end{figure}

Let us now focus on the short scale behaviour. We see that the divergent behaviour of the
graviton correlations gets regularized at high energies by the discrete structure of spin foams.
This is how the full theory enhances the effective field theory where the latter breaks down,
and it confirms the intuition that spacetime can not be considered as fluctuating around the flat metric 
at the Planck scale. 
The peak 
is very close to the Planck scale, and its exact location depends
on the value of $\alpha$, but also on the measure term in \Ref{asymp}. 
To show this, in Fig.~\ref{kfig} we plot again the case $\alpha=5$, but this time with $k=1$.
Notice that the position of the peak is slightly pushed to the right, but the qualitative behaviour is the same. 

We want to conclude with an outlook for further developments to turn this picture into a 
concrete prediction of the theory. This will require enhancing both the boundary state and the kernel
used here.

The first improvement would be to remove the ambiguity in $\alpha_{lm}$ by fixing it with a dynamical requirement.
To obtain a quantitative matching with the free graviton $G_{ab}$ we need the right boundary state, with the right
non-diagonal structure and values of the entries, coming from the dynamics and a choice of gauge.
As mentioned above, a possible way to obtain this quantity is to evaluate it in linearized area Regge calculus.

Furthermore, from analogies with conventional quantum field theory we expect there to exist
a procedure for extracting the full boundary state from the non-perturbative kernel. 
Extending this procedure in background independent quantum gravity (see for instance \cite{PhysBS}) 
would be extremely useful.
The full state will fix $\alpha_{lm}$ by looking at the perturbative expansion, but also contribute to the small scale
structure, thus affecting the numerical analysis presented here. 
This will certainly modify numerical results such as the exact location of the peak, but will most likely 
not change the qualitative picture of suppressed correlations at the Planck scale.
This regularizing effect is very likely to survive in the full
theory, because it is a facet of the discrete microscopic structure, more than of the details of the model.

On the other hand, the kernel itself needs improvement.
Indeed the BC kernel does not capture the
dynamics of quantum gravity in a fully consistent way. Here we used it in the restricted
context of area correlations on a single 4-simplex, where it provides a sensible
quantum gravity amplitude.
To test the full tensorial structure of \Ref{Wlin}, we need to consider
also projections that in the discrete picture would correspond to correlations between the dihedral angles
of the boundary geometry. As pointed out in \cite{grav2,Alesci}, these correlations can not be studied using the
BC vertex amplitude.
Furthermore, the calculations presented here need to be extended to many 4-simplices, and the large spin
limit of BC \Ref{asymp} precisely lacks the constraints which are needed to correctly treat the areas
as independent variables.
These issues have recently been addressed, and promising new models have been proposed \cite{newmodels}.
It would be extremely interesting to apply the same analysis to these new models.

In conclusion, we have presented a spin foam model where
correlations consistent with the graviton theory at low energies
are dynamically suppressed and regularized at the Planck scale.
This work shows that the spin foam framework for quantum gravity naturally provides the expected regularization 
of the corrections at short scales, i.e.\ high energies: the theory has a short length scale
appearing dynamically, which also suggests that spacetime can not be considered as fluctuating around 
the flat metric at the Planck scale.
We stress that the model presented here is not predictive, and further work is needed before
the correctness of the semiclassical limit of spin foam gravity can be claimed. 
Our results give a glimpse of what the qualitative picture of quantum gravity and its bridging
to low-energy physics could be like, and suggest interesting new questions for further
investigation.

\end{document}